\documentclass[
reprint,
superscriptaddress,
amsmath,
amssymb,
aps,
prd
]{revtex4-1}

\usepackage{graphicx}
\usepackage{dcolumn}
\usepackage{bm}
\usepackage{amssymb}   
\usepackage{hyperref}
\usepackage{aas_macros}

\newcommand{\HI}{H\,{\sc i}}

\begin{document}

\preprint{APS/123-QED}

\title{Detecting Cosmic Reionization using Bi-Spectrum Phase}

\author{Nithyanandan Thyagarajan}
\email{t\_nithyanandan@nrao.edu}
\homepage{https://tnithyanandan.wordpress.com/}
\thanks{Nithyanandan Thyagarajan is a Jansky Fellow of the National Radio Astronomy Observatory.}
\affiliation{National Radio Astronomy Observatory, Socorro, NM 87801, USA}
\affiliation{Arizona State University, School of Earth and Space Exploration, Tempe, AZ 85287, USA}
\author{Chris L. Carilli}%
\affiliation{National Radio Astronomy Observatory, Socorro, NM 87801, USA}
\affiliation{Astrophysics Group, Cavendish Laboratory, University of Cambridge, Cambridge CB3 0HE, UK}


\author{Bojan Nikolic}
\affiliation{Astrophysics Group, Cavendish Laboratory, University of Cambridge, Cambridge CB3 0HE, UK}%


\date{\today}

\begin{abstract} 
  Detecting neutral Hydrogen (H\,{\sc i}) via the 21~cm line emission from the intergalactic medium at $z\gtrsim 6$ has been identified as one of the most promising probes of the epoch of cosmic reionization -- a major phase transition of the Universe. However, these studies face severe challenges imposed by the bright foreground emission from cosmic objects. Current techniques require precise instrumental calibration to separate the weak H\,{\sc i} line signal from the foreground continuum emission. We propose to mitigate this calibration requirement by using measurements of the interferometric bi-spectrum phase. Bi-spectrum phase is unaffected by antenna-based direction-independent calibration errors and hence for a compact array it depends on the sky brightness distribution only (subject to the usual thermal-like noise). We show that the bi-spectrum phase of foreground synchrotron continuum has a characteristically smooth spectrum relative to the cosmological line signal. The two can be separated effectively by exploiting this spectral difference using Fourier techniques, while eliminating the need for precise antenna-based calibration of phases introduced by the instrument, and the ionosphere, inherent in existing approaches. Using fiducial models for continuum foregrounds, and for the cosmological H\,{\sc i} signal, we show the latter should be detectable in bi-spectrum phase spectra, with reasonable significance at $|k_\parallel| \gtrsim 0.5\,h$~Mpc$^{-1}$, using existing instruments. Our approach will also benefit other H\,{\sc i} intensity mapping experiments that face similar challenges, such as those measuring Baryon Acoustic Oscillations (BAO).
\end{abstract}




\maketitle


\section{Introduction}\label{sec:intro}

The hydrogen gas that dominates the baryon content of the Universe has undergone two phase transitions over cosmic history. The first was the transition from fully ionized to fully neutral gas during cosmic \emph{recombination\/} $\sim 300,000$ years after the Big Bang ($z\approx 1100$). This epoch has been quantified in exquisite detail through cosmic microwave background radiation studies \cite{planck15i}. The second transition is known as cosmic \emph{reionization}, when light from the first stars and black holes reionized the neutral Hydrogen (\HI) gas that pervaded the early Universe. Current indirect constraints suggest this second transition occurred a few hundred Myr to $\sim 1$~Gyr ($z\gtrsim 6$) \cite{gre17a} after the Big Bang. During this epoch, the photon-to-baryon ratio rose significantly, reionizing the entire Universe except gas bound to galaxies. The astrophysical processes during this epoch shaped the evolution of galaxies and large-scale structure. However, unlike recombination, the timing and process of cosmic reionization remain poorly understood. The redshifted 21~cm spectral line from the electron spin-flip transition in \HI\ atom, has been recognized as the most promising tool to unravel the physical processes involved in cosmic reionization, and the evolution of large scale structure during the formation of the first galaxies \cite{sun72,sco90,mad97,toz00,ili02,fan02,fan06,bar07,mor10}.

Our Galaxy and intervening radio galaxies emit synchrotron radiation in the same frequency band ($\sim$~100--200~MHz) as the redshifted \HI\ signal from the epoch of reionization (EoR), with a brightness temperature roughly $\gtrsim 10^4$ times higher than the EoR \HI\ signal. However, the synchrotron foregrounds have a smooth spectrum whereas the EoR \HI\ signal will appear as small fluctuations superimposed on the smooth foreground spectrum. The hope of separating the cosmic line signal from the Galactic and extragalactic foregrounds has spawned a number of low-frequency instruments (eg. \cite{par10,tin13,van13}), with sensitivities sufficient for a statistical detection of the EoR \HI\ signal using power spectrum methodology \cite{thy13,bea13}.  However, despite obtaining significant observational data, these experiments remain dynamic range limited due to the coupling of the strong foreground emission to the EoR \HI\ signal via instrumental effects, particularly systematics related to precision of calibration, and the intrinsic chromatic instrumental response \cite{thy15a,thy15b,thy16,dat09,dat10,tro16, pac13,ali15,patil17,pob15,liu10,zhe14,barry16,sie17,dil17}. 

We present an alternate approach using the phase of the complex bi-spectrum (also referred to in radio interferometry as closure phase) to detect the signal from the EoR. This quantity is impervious to antenna-based complex gains introduced by the instrument and the ionosphere, and thus can remove the need for high-precision calibration currently impeding existing approaches\cite{car16}. 

Further, \HI\ intensity mapping experiments, such as CHIME \cite{ban14}, HIRAX \cite{new16}, and Tianlai \cite{xu15} will map \HI\ on large-scales at $1\lesssim z \lesssim 3$ tracing the matter density fluctuations, and thus measure the Baryon Acoustic Oscillations (BAO) that can be used as a standard ruler in constraining the {\it Dark Energy} equation of state. The challenges in these experiments are expected to be of a similar magnitude to that in 21~cm EoR experiments, and thus can potentially benefit from our approach.

\section{Properties of Bi-spectrum Phase}\label{sec:CPinfo}

The bi-spectrum in the context of interferometry has been investigated in \cite{jen58,kul89,tay99,tho01,mon06} and recently revisited in \cite{car18}. Consider a triad of antennas, $\{a, b, c\}$, indexed by $i$. $V_{ij}^\textrm{m}(f)$ is the spatial coherence measured at frequency, $f$, between antenna pairs, $\{ab, bc, ca\}$, indexed by $ij$. An interferometer measures $V_{ij}^\textrm{m}(f) = g_i(f)\, g_j^*(f)\, V_{ij}^\textrm{T}(f) + V_{ij}^\textrm{N}(f)$, the  true spatial coherence, $V_{ij}^\textrm{T}(f)$, corrupted by multiplicative antenna gains and ionospheric distortions, $g_i$, and additive thermal-like noise, $V_{ij}^\textrm{N}(f)$. We express $V_{ij}^\textrm{T}(f) = V_{ij}^\textrm{F}(f) + V_{ij}^\textrm{H{\sc i}}(f)$, where, $V_{ij}^\textrm{F}(f)$ and $V_{ij}^\textrm{H{\sc i}}(f)$ are due to foreground synchrotron and EoR \HI\ signal, respectively. The measured bi-spectrum is $B_\nabla^\textrm{m}=\prod_{ij} V_{ij}^\textrm{m}=B_\nabla^\textrm{T}\prod_i |g_i|^2+B_\nabla^\textrm{N}$, where, the dependence on $f$ has been omitted for convenience (hereafter so, unless indicated). In our notation, $ij$ only indexes three specific cyclic antenna pairs $\{ab, bc, ca\}$ and not all possible combinations of $i$ and $j$. $B_\nabla^\textrm{N}$, encompassing all terms containing $V_{ij}^\textrm{N}$ including cross-terms, has zero mean since it contains $V_{ij}^\textrm{N}$, which is assumed to be uncorrelated between antenna pairs. The phase of $B_\nabla^\textrm{m}$ is:
\begin{align}
  \phi_\nabla^\textrm{m} &= \sum_{ij}\phi_{ij}^\textrm{m} = \phi_\nabla^\textrm{T} + \delta\phi_\nabla^\textrm{N} = \left(\phi_\nabla^\textrm{F} + \delta\phi_\nabla^\textrm{H{\sc i}}\right) + \delta\phi_\nabla^\textrm{N} \label{eqn:cpphase-sum},
\end{align}
where, $\phi_{ij}^\textrm{m} = \phi_i - \phi_j + \phi_{ij}^\textrm{T} + \delta\phi_{ij}^\textrm{N}$, and $\phi_i$ are the phases of $V_{ij}^\textrm{m}$ and $g_i$, respectively. $\delta\phi_\nabla^\textrm{H{\sc i}}$ and $\delta\phi_\nabla^\textrm{N}$ are not just determined by the phases of $V_{ij}^\textrm{H{\sc i}}$ and $V_{ij}^\textrm{N}$ but are small  perturbations due to the EoR \HI\ signal and noise respectively, superimposed on dominant phase, $\phi_\nabla^\textrm{F}$, due to foregrounds, and depend on $\rho_{ij}^\textrm{H{\sc i}}\equiv |V_{ij}^\textrm{F}|/|V_{ij}^\textrm{H{\sc i}}|$ and $\rho_{ij}^\textrm{N}\equiv |V_{ij}^\textrm{F}|/|V_{ij}^\textrm{N}|$ respectively. The small-perturbation formalism is valid when $\rho_{ij}^\textrm{H{\sc i}}\gg 1$ and $\rho_{ij}^\textrm{N}\gg 1$, as is usually the case in low-frequency cosmology experiments. $\phi_\nabla^\textrm{m}$ is independent of the antenna gains and measures $\phi_\nabla^\textrm{T}$, corrupted only by noise, $\delta\phi_\nabla^\textrm{N}$. Hence, $\phi_\nabla^\textrm{m}$ contains information purely about the sky's spatial structure, with certain geometric properties \cite[see][]{mon07}.

$\delta\phi_{ij}^\textrm{N}$ follows a Gaussian distribution with standard deviation, $\sigma_{\phi_{ij}^\textrm{N}} = (\sqrt{2}\,\rho_{ij}^\textrm{N})^{-1}$, when $\rho_{ij}^\textrm{N}\gg 1$ \cite{cra89}. From Eq.~(\ref{eqn:cpphase-sum}), for $\rho_{ij}^\textrm{N}\gg 1$, $\phi_\nabla^\textrm{m}$ will also approach a Gaussian distribution with variance, $\sigma_{\phi_\nabla^\textrm{m}}^2 = \sum_{ij}\,(\sqrt{2}\,\rho_{ij})^{-2}$, where, $\rho_{ij}=\rho_{ij}^\textrm{N}$ (if $\rho_{ij}^\textrm{N} < \rho_{ij}^\textrm{H{\sc i}}$) or $\rho_{ij}=\rho_{ij}^\textrm{H{\sc i}}$ (if $\rho_{ij}^\textrm{H{\sc i}} < \rho_{ij}^\textrm{N}$). $\delta\phi_\nabla^\textrm{H{\sc i}}$ can be made the dominant fluctuation in $\phi_\nabla^\textrm{m}$ by averaging a number of independent samples ($N_\textrm{m}$) of $\phi_\nabla^\textrm{m}$ and reducing the variance of $\delta\phi_\nabla^\textrm{N}$ to $\sigma_{\delta\phi_\nabla^\textrm{N}}^2 / N_\textrm{m}$. 

\section{Modeling}\label{sec:modeling}

\subsection{Instrument Model}\label{sec:instrument}

We use the Hydrogen Epoch of Reionization Array \cite[HERA;][]{deb17,thy16,ewa16,neb16,patra17} as an example to demonstrate our approach. Located at the Karoo desert in South Africa at a latitude of $-30.72^\circ$, HERA will consist of 350 close-packed 14~m dishes with a shortest antenna spacing of 14.6~m. Its triple split-core hexagonal layout is optimized for both redundant calibration and the delay spectrum technique for detecting the cosmic EoR \HI\ signal \cite{dil16}. 

\subsection{Sky Model and Noise}\label{sec:skymodel-noise}

We construct an all-sky model that includes the realization of a fiducial {\sc faint galaxies} EoR model \cite{gre17b} using 21cmFAST \cite{mes11} and a radio foreground model that includes compact and diffuse synchrotron emission from the Galaxy and extragalactic sources as in \cite{thy15a}. 

\begin{figure}[htb]
\includegraphics[width=\linewidth]{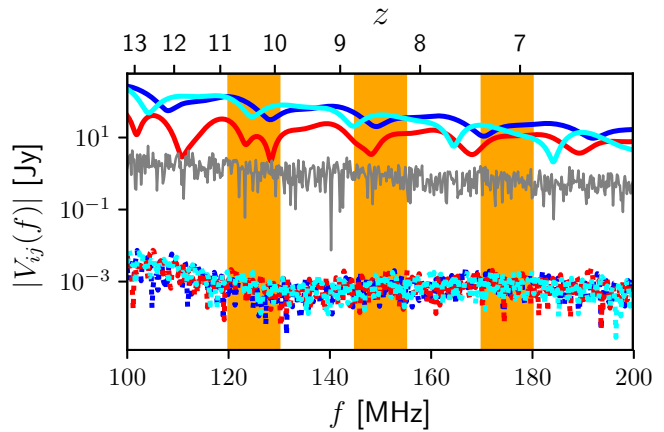}
\caption{Spatial coherence amplitude spectra on antenna pairs forming a 14.6~m equilateral triad are shown in red, blue and cyan. Solid and dotted lines show amplitudes from synchrotron foregrounds and a fiducial EoR model, respectively. Gray curve shows the typical noise obtained with 1~min integration. Typically, the foregrounds are $\gtrsim 10^4$ times brighter than the EoR signal. The orange bands denote the frequency sub-bands centered on 125~MHz, 150~MHz, and 175~MHz each of 10~MHz effective bandwidth. \label{fig:vis-spectra}}
\end{figure}

Fig.~\ref{fig:vis-spectra} shows $|V_{ij}(f)|$ measured on three non-redundant 14.6~m antenna spacings due to the sky (foreground synchrotron and EoR \HI) and thermal noise contributions with 1~min integration. The foreground contributions exceed the EoR signal by a factor $\sim 10^4$ as expected. It is also noted that $\rho_{ij}^\textrm{N} \gtrsim 100$, which implies $\delta\phi_\nabla^\textrm{N}$ will follow a Gaussian distribution with variance, $\sigma_{\delta\phi_\nabla^\textrm{N}}^2$.

\begin{figure}[htb]
\includegraphics[width=\linewidth]{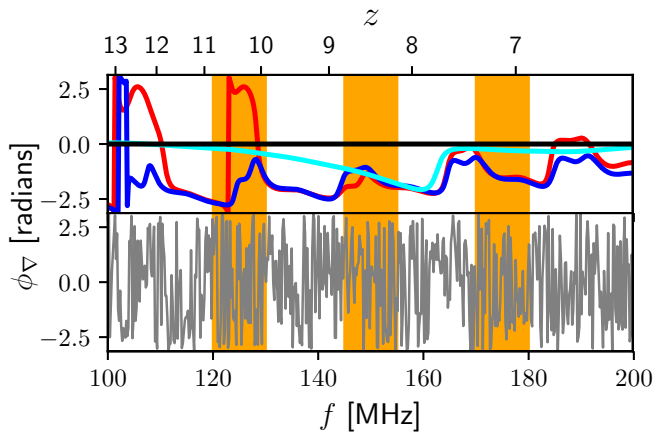}
\caption{Bi-spectrum phase spectra of individual components in the sky model -- single point source (black) with $\phi_\nabla=0$, compact sources (cyan), diffuse foregrounds (blue), diffuse and compact components combined (red), and the EoR \HI\ fluctuations (gray) -- for a 14.6~m equilateral triad. The \HI\ component is highly fluctuating relative to the foregrounds. The phase wrap discontinuities at $\phi_\nabla=\pm \pi$ are not of a physical origin. The orange sub-bands are same as in Fig.~\ref{fig:vis-spectra}. \label{fig:cp-spectra}}
\end{figure}

Fig.~\ref{fig:cp-spectra} shows the the bi-spectrum phase spectra for the foreground and EoR components separately. Clearly, the foreground component is characterized by a smoother spectrum whereas the EoR component fluctuates rapidly. The observed $\phi_\nabla^\textrm{m}$, that includes the foregrounds, EoR \HI\, and noise components, will be described by Eq.~\ref{eqn:cpphase-sum}. 

\section{Extraction of the cosmic signal}\label{sec:extraction}

The contrasting spectral characteristics -- smooth $\phi_\nabla^\textrm{F}$ and fluctuating $\delta\phi_\nabla^\textrm{H{\sc i}}$ -- indicate that techniques similar to the power spectrum approaches could be employed to separate the cosmic signal from foregrounds but avoiding the need for high-precision spectral calibration that most existing approaches rely on.

If a number of independent measurements are available, they could be used to average $\phi_\nabla^\textrm{m}$ to reduce the standard deviation of $\delta\phi_\nabla^\textrm{N}$ by taking advantage of its Gaussian distribution. When $\delta\phi_\nabla^\textrm{N}$ is sufficiently low, $\delta\phi_\nabla^\textrm{H{\sc i}}$ will dominate the spectral fluctuations in $\phi_\nabla^\textrm{m}$, while $\phi_\nabla^\textrm{F}$ will be dominant overall but spectrally smooth. The variance in $\delta\phi_\nabla^\textrm{H{\sc i}}$ is effectively a measure of the EoR \HI\ line strength to foreground continuum ratio.

Thus, when $\rho_{ij}^\textrm{H{\sc i}} < \rho_{ij}^\textrm{N}$, $\sigma_{\phi_\nabla^\textrm{m}}$ can be used to estimate $\rho_{ij}^\textrm{H{\sc i}}$, which in turn can be used to infer $|V_{ij}^\textrm{H{\sc i}}|$ if $|V_{ij}^\textrm{F}|$ is known. $|V_{ij}^\textrm{H{\sc i}}|$ is directly related to the EoR H\,{\sc i} brightness temperature fluctuations, $\langle(\delta T_\textrm{b}^\textrm{H{\sc i}})^2\rangle$, which can be inferred as a function of  redshift. This is discussed further below.

\subsection{Delay Power Spectrum of Bi-spectrum Phase}\label{sec:cp-FG-wedge}

While any method that relies on separating a signal from contaminants using differences in spectral characteristics is applicable, we employ the {\it Delay Spectrum} approach \cite{par12a,par12b}. Since the $\phi_\nabla$ spectrum can have discontinuities at $\phi_\nabla=\pm\pi$, and hence non-physical spectral features, we define $\xi_\nabla = e^{i\phi_\nabla^\textrm{m}}$, which is unaffected by these discontinuities. We perform a delay transform:
\begin{align}\label{eqn:cpdspec}
  \Xi_\nabla(\tau) &= \int \xi_\nabla(f)\,W(f)\,e^{i2\pi f\tau}\,\mathrm{d}f,
\end{align}
where, $W(f)$ is a spectral weighting usually chosen to control the quality of the delay spectrum \citep{thy13,thy16} and has an effective bandwidth, $B_\textrm{eff}$. $\Xi_\nabla(\tau)$ has units of Hz. 

We define the delay cross-power spectrum of $\xi_\nabla(f)$ as:
\begin{align}
  P_\nabla(k_\parallel) &\equiv \mathfrak{Re}\bigg\{\Xi_\nabla(\tau)\,\Xi_{\nabla^\prime}^*(\tau)\bigg\} \left(\frac{\Delta D}{B_\textrm{eff}^2}\right), \label{eqn:delay-power-spectrum}
\end{align}
where, $\Delta D\equiv \Delta D(z)$ is the comoving depth along the line of sight corresponding to $B_\textrm{eff}$ at redshift $z$, $k_\parallel\approx 2\pi\tau B_\textrm{eff}/\Delta D$ is line-of-sight wavenumber, and $\mathfrak{Re}\{\cdot\}$ denotes the real part. In this paper, we use cosmological parameters from \cite{planck15xiii} with $H_0=100\,h$~km~s$^{-1}$~Mpc$^{-1}$. $P_\nabla(k_\parallel)$ is in units of $h^{-1}$~Mpc. When $\Xi_\nabla(\tau)=\Xi_{\nabla^\prime}(\tau)$, $P_\nabla(k_\parallel)$ reduces to delay auto-power spectrum. In \S\ref{sec:averaging}, we describe the usage of Eq.~(\ref{eqn:delay-power-spectrum}) in different contexts. 

From Eq.~(\ref{eqn:cpphase-sum}), since $\xi_\nabla = \prod_{ij} e^{i\phi_{ij}^\textrm{m}}$, it can be shown that $\Xi_\nabla(\tau) = \Xi_{ab}(\tau)\circledast \Xi_{bc}(\tau)\circledast \Xi_{ca}(\tau)\circledast \mathcal{W}(\tau)$, where, $\{\xi_{ij}(f),\,\Xi_{ij}(\tau)\}$, and $\{W(f),\,\mathcal{W}(\tau)\}$ are Fourier transform pairs, and $\circledast$ denotes convolution. Foreground contribution to spatial coherence is known to be confined to delay modes within the {\it horizon limits} \cite{bow09,liu09,liu14a,liu14b,dat10,liu11,gho12,mor12,par12b,tro12,ved12,dil13,pob13,thy13,thy15a,thy15b,thy16,dil14}. As a result of the convolution, the foreground contribution to $\Xi_\nabla(\tau)$ is expected to be more extended along $\tau$ (and $k_\parallel$) relative to $\Xi_{ij}(\tau)$. This relation will be explored in a forthcoming study. 

\subsection{Improving sensitivity}\label{sec:averaging}

We require $\sigma_{\delta\phi_\nabla^\textrm{N}} < \sigma_{\delta\phi_\nabla^\textrm{H{\sc i}}}$ such that $\phi_\nabla^\textrm{F}, \delta\phi_\nabla^\textrm{H{\sc i}} > \delta\phi_\nabla^\textrm{N}$ in order to detect EoR and estimate $\rho_{ab}^\textrm{H{\sc i}}$. This can be achieved by a combination of the following.

First, $\xi_\nabla$ can be averaged coherently until the variation due to the transiting sky exceeds the decrease in noise due to averaging \cite{car18}. An Allan Variance estimate for HERA showed this timescale to be $\lesssim 2$~min \cite{car18}.

Second, $\xi_\nabla$ can be averaged coherently across multiple days at a fixed LST before computing $P_\nabla(k_\parallel)$, as long as the array size is small compared to the predominant spatial scales in the ionospheric variations, as has been observed to be the case for HERA \cite{car18}. 

Third, the measurements on redundant triads can be averaged coherently in $\xi_\nabla$, before computing the power spectrum.  For non-redundant triads (including different triad classes -- equilateral, isosceles, and scalene), it may be possible to average incoherently in $P_\nabla(k_\parallel)$, using  $\Xi_\nabla(\tau)$ and $\Xi_{\nabla^\prime}^*(\tau)$ from triad pairs $\nabla$ and $\nabla^\prime$ respectively to further improve sensitivity. 

Lastly, since the EoR signal is statistically isotropic in space while the foregrounds are not, measurements on time intervals larger than the coherence timescale can be used to compute the individual delay cross-power spectra, $P_\nabla(k_\parallel)$, from temporal pairs $\Xi_\nabla(\tau)$ and $\Xi_{\nabla^\prime}^*(\tau)$, and then be averaged incoherently. 

\subsection{Detection of Cosmic Reionization}\label{sec:EoR-detection}

The primary application of this technique is to detect the EoR \HI\ signal. Using an {\it Airy} power pattern for the HERA dish, we used PRISim \footnote{The Precision Radio Interferometry Simulator (PRISim) Python package is available at \href{https://github.com/nithyanandan/PRISim}{https://github.com/nithyanandan/PRISim}.} to simulate $V_{ij}^\textrm{F}(f)$, $V_{ij}^\textrm{H{\sc i}}(f)$, and $V_{ij}^\textrm{N}(f)$ from the synchrotron foregrounds, an EoR \HI\ model, and noise, respectively as described in \S\ref{sec:modeling} over the 100--200~MHz band with 195.3125~kHz spectral resolution, and a conservative phase-coherent integration interval of 1~min \cite{car18}. 

For bi-spectrum phase, we consider 14.6~m equilateral antenna triads and assume ideal redundancy for the array. The total number of such measurements is assumed to be $N_\textrm{m} \sim 10^6$ (after allowing for $\sim$~50\% efficiency of data quality from $\sim 30$ redundant 14.6~m equilateral triads with HERA-47 \cite[see layout in][]{car18}, $\sim$~8~hours of observing per day at 1~min integration intervals, repeated over $\sim$~150 nights). We assume there are $N_\textrm{c}$ coherent measurements of $\phi_\nabla^\textrm{m}$ and for each of these measurements, there are $N_\textrm{ic}$ incoherent measurements such that $N_\textrm{c}N_\textrm{ic}=N_\textrm{m}$. $\phi_\nabla^\textrm{m}$ is averaged coherently over $N_\textrm{c}$ measurements, and then averaged incoherently in $P_\nabla(k_\parallel)$ measured for all incoherent pairs of $\phi_\nabla^\textrm{m}$. We choose $W(f)$ to be the inverse Fourier transform of the squared delay response of a {\it Blackman-Harris} spectral weighting \cite{har78} as proposed in \cite{thy16}, on sub-bands centered at 125~MHz, 150~MHz, and 175~MHz, with an effective bandwidth of $B_\textrm{eff}\simeq 10$~MHz (to minimize EoR signal evolution over the redshift range).

\begin{figure*}[htb]
  \includegraphics[width=\linewidth]{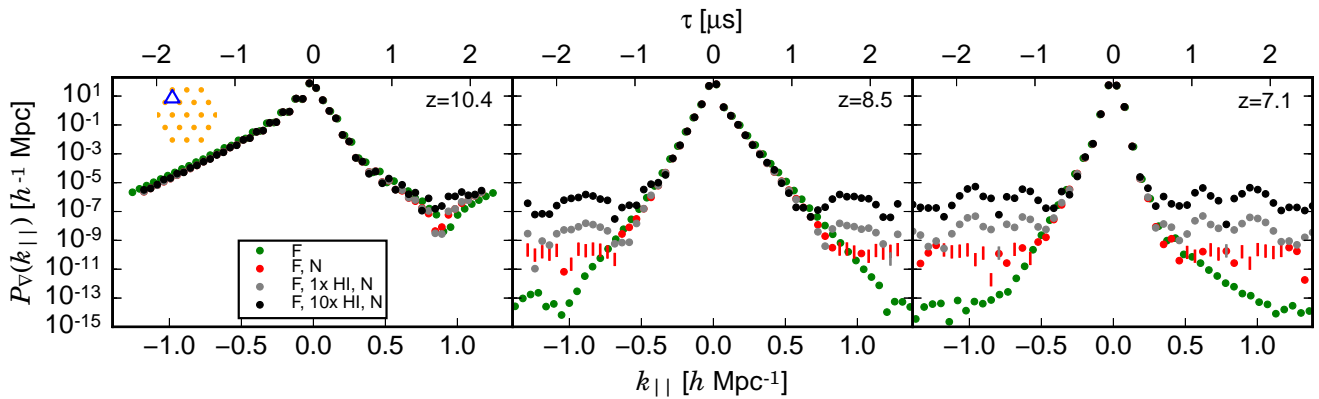}
\caption{Delay power spectra of bi-spectrum phases in sub-bands centered at 125~MHz (left), 150~MHz (middle), and 175~MHz (right) that include contributions from foregrounds, EoR signal, and noise. The corresponding redshifts are listed in each panel. The 14.6~m equilateral triad used is highlighted on a HERA-19 layout on the top left. Green curves show only the foreground contribution ($\phi_\nabla^\textrm{m}=\phi_\nabla^\textrm{F}$). The anisotropic foreground model employed \cite{thy15a} results in the asymmetry around $k_\parallel=0\,h^{-1}$~Mpc. The red curves show the effect of including noise fluctuations ($\phi_\nabla^\textrm{m}=\phi_\nabla^\textrm{F}+\delta\phi_\nabla^\textrm{N}$), after averaging $N_\textrm{m} \sim 10^6$ measurements. The gray curves apply when fluctuations from the EoR \HI\ signal are included, $\phi_\nabla^\textrm{m}=\phi_\nabla^\textrm{F}+\delta\phi_\nabla^\textrm{H{\sc i}}+\delta\phi_\nabla^\textrm{N}$. The black curves are identical to the gray ones except the EoR \HI\ signal is 10 times brighter. The dot and vertical markers denote positive and negative values respectively. The latter results from the removal of noise bias by using cross-power spectra (Eq.~\ref{eqn:delay-power-spectrum}). The EoR \HI\ contribution is significantly detected at $|k_\parallel| \gtrsim 0.5\,h$~Mpc$^{-1}$ at $z\approx 8.5$ and $z\approx 7.1$. At $z\approx 10.4$, the EoR contribution is inseparable from foregrounds because the EoR signal is weaker (see power spectrum of {\sc faint galaxies} model at $k=0.5$~Mpc$^{-1}$ in Fig.~2 of \cite{gre17b}) while the foregrounds are brighter. The detection significance depends on the strength of EoR H\,{\sc i} signal relative to the foregrounds. A 10-fold increase in EoR intensity shows a $\sim 100$-fold increase in $P_\nabla(k_\parallel)$ at $|k_\parallel| \gtrsim 0.5\,h$~Mpc$^{-1}$ at $z\approx 8.5$ and $z\approx 7.1$, indicating its sensitive dependence on $\rho_{ij}^\textrm{H{\sc i}}$. This detection measures the line-to-continuum ratio, from which the EoR H\,{\sc i} brightness temperature fluctuations, $\langle(\delta T_\textrm{b}^\textrm{H{\sc i}}(z))^2\rangle$, can be inferred using a reliable foreground model. \label{fig:cpdps}}
\end{figure*}

Fig.~\ref{fig:cpdps} shows the delay cross-power spectra, $P_\nabla(k_\parallel)$, of $\phi_\nabla^\textrm{m}$ obtained for the three sub-bands, with an initial S/N, $\rho_{ij}^\textrm{N}=200$. It is seen that the contributions from the EoR \HI\ fluctuations, $\delta\phi_\nabla^\textrm{H{\sc i}}$, are detected as they dominate over $\phi_\nabla^\textrm{F} + \delta\phi_\nabla^\textrm{N}$ at $|k_\parallel| \gtrsim 0.5\,h$~Mpc$^{-1}$ in the 150~MHz and 175~MHz sub-bands. The absence of a clear detection in the 125~MHz sub-band is discussed below.

\subsection{EoR H\,{\sc i} Brightness Temperature Fluctuations} \label{sec:Tb-variance}

Beyond simple detection, we could also estimate the EoR \HI\ brightness temperature fluctuations, $\langle(\delta T_\textrm{b}^\textrm{H{\sc i}}(z))^2\rangle$, at various redshifts using the separation in $P_\nabla(k_\parallel)$. When $\rho_{ij}^\textrm{H{\sc i}} < \rho_{ij}^\textrm{N}$, the separation between $\phi_\nabla^\textrm{m}$ and $\phi_\nabla^\textrm{F} + \delta\phi_\nabla^\textrm{N}$ in $P_\nabla(k_\parallel)$ depends sensitively on $\rho_{ab}^\textrm{H{\sc i}}$. Fig.~\ref{fig:cpdps} shows $P_\nabla(k_\parallel)$ in the three sub-bands for different values of $\langle(\delta T_\textrm{b}^\textrm{H{\sc i}}(z))^2\rangle$, or equivalently, $\rho_{ij}^\textrm{H{\sc i}}$. As the EoR \HI\ signal strength increases (ten-fold in $\langle(\delta T_\textrm{b}^\textrm{H{\sc i}}(z))^2\rangle^{1/2}$), the separation in $P_\nabla(k_\parallel)$ also increases at $|k_\parallel| \gtrsim 0.5\,h$~Mpc$^{-1}$ ($\sim 100$ times), most clearly in the 150~MHz and 175~MHz sub-bands. Conversely, this separability can be used to estimate $\rho_{ij}^\textrm{H{\sc i}}$ in the different sub-bands, which in turn can be used to estimate $\langle(\delta T_\textrm{b}^\textrm{H{\sc i}}(z))^2\rangle$ by using the best foreground models available in those sub-bands. The accuracy of $\langle(\delta T_\textrm{b}^\textrm{H{\sc i}}(z))^2\rangle$ so determined will depend on the uncertainty in the foreground model, which is not required to be as precise as the calibration requirement.

The redshift evolution of separability in $P_\nabla(k_\parallel)$ depends on relative strengths of the foreground and the EoR \HI\ signal. In the 175~MHz sub-band ($z\approx 7.1$), relative to the 150~MHz sub-band ($z\approx 8.5$), $\langle(\delta T_\textrm{b}^\textrm{H{\sc i}}(z))^2\rangle$ is weaker (see {\sc faint galaxies} model at $k=0.5$~Mpc$^{-1}$ in Fig.~2 of \cite{gre17b}) but the foreground synchrotron temperature is also fainter roughly by the same factor. Therefore, the separability of EoR \HI\ in $P_\nabla(k_\parallel)$ is roughly similar in the two sub-bands. However, in the 125~MHz sub-band ($z\approx 10.4$), $\langle(\delta T_\textrm{b}^\textrm{H{\sc i}}(z))^2\rangle$ in the fiducial EoR model is fainter but the foregrounds are brighter, thereby significantly decreasing the spectral contrast between foregrounds and the EoR signal. Hence, a clear separability is not seen at $z\approx 10.4$.

This hypothesis can be confirmed by estimating $\rho_{ij}^\textrm{H{\sc i}}$ in different sub-bands and verifying that there is no separation in $P_\nabla(k_\parallel)$ at any $k$ at $z\lesssim 6$, when the IGM is understood to be fully ionized, with a reasoning similar to \cite{pob16b}. Lack of separation in $P_\nabla(k_\parallel)$, such as at $z\approx 10.4$, will enable placing upper limits on $\langle (\delta T_\textrm{b}^\textrm{H{\sc i}}(z))^2\rangle$.

\section{Summary}\label{sec:summary}

A primary limitation to existing low-frequency EoR 21~cm cosmology experiments remains high-precision spectral calibration in order to remove the strong continuum foregrounds. We propose a new approach that uses bi-spectrum phase, which represents an intrinsic property of the sky and is thus impervious to antenna-based calibration and associated errors, unlike existing approaches.
  
The primary goal of the proposed technique is to obtain an interferometric detection of the \HI\ 21~cm signal from the neutral IGM during cosmic reionization. The next goal will be to relate the measured EoR \HI\ bi-spectrum phase power spectrum to the astrophysical quantities of interest, particularly, the brightness temperature fluctuations. The magnitude of the line signal relative to that of the foreground continuum in the bi-spectrum phase power spectrum, is essentially a measure of the line-to-continuum ratio, since the line signal introduces small, frequency-dependent, perturbations on the spatial coherence phase. Hence, the line signal strength can be calibrated against a foreground model.

Ultimately, the quantitative interpretation will entail a standard forward-modeling process, comparing measurements to the predictions of bi-spectrum phase power spectra for different sky models. This work investigates one such realization, and demonstrates that a quantitative relationship exists between the EoR signal strength and the bi-spectrum phase power spectra.

Beyond the simple demonstration presented here, there are many ways to improve the prospects of detecting the cosmic reionization signal using the bi-spectrum phase as outlined in \S\ref{sec:averaging} using optimal weighting techniques \cite{liu14a,liu14b,dil15}, particularly in the context of a redundant array such as HERA (though it applies to non-redundant arrays as well). Upon completion, HERA will consist of 350 antennas yielding $\sim$~10--100 times more measurements than used here, and thereby go deeper in overall sensitivity. Removal of the foreground contribution in $\xi_\nabla$ could potentially relax the dynamic range required for detection and boost sensitivity in $|k_\parallel| \lesssim 0.5\,h$~Mpc$^{-1}$. 

\begin{acknowledgments}
We acknowledge the insightful and helpful inputs from Rajaram Nityananda, Gianni Bernardi, Judd Bowman, James Aguirre, Aaron Parsons, James Moran, Jonathan Pober, Adam Beardsley, and Peter Williams that helped in the preparation of this manuscript. The National Radio Astronomy Observatory is a facility of the National Science Foundation operated under cooperative agreement by Associated Universities, Inc. We acknowledge support from the Royal Society and the Newton Fund under grant NA150184.
\end{acknowledgments}


%

\end{document}